\documentstyle[12pt,twoside,fleqn,espcrc1,axodraw,epsf]{article}

\newcommand{\be}{\begin{equation}}
\newcommand{\ee}{\end{equation}}
\newcommand{\ba}{\begin{eqnarray}}
\newcommand{\ea}{\end{eqnarray}}
\newcommand{\rmi}[1]{{\mbox{\scriptsize #1}}}

\newcommand{\nn}{\nonumber}

\newcommand{\fm}{\rm fm}
\newcommand{\gev}{\rm GeV}
\newcommand{\psat}{p_\rmi{sat}}

\newcommand{\fr}[2]{{\frac{#1}{#2}}}

\def\lsim{\raise0.3ex\hbox{$<$\kern-0.75em\raise-1.1ex\hbox{$\sim$}}}
\def\gsim{\raise0.3ex\hbox{$>$\kern-0.75em\raise-1.1ex\hbox{$\sim$}}}

\newcommand{\AmS}{{\protect\the\textfont2
  A\kern-.1667em\lower.5ex\hbox{M}\kern-.125emS}}

\title{From QCD to heavy ion collisions}

\author{K. Kajantie\address{Department of Physics, P. O. Box 9,
FIN-00014 University of Helsinki, Finland}}

\begin{document}
\maketitle

\begin{abstract}
This talk will discuss how heavy ion experiments, when moving
from SPS (10 + 10 GeV) to RHIC (100+100 GeV)
and to LHC (2750+2750 GeV), will enter a new domain of QCD in which
the production of even large $p_T$ gluons is so abundant that it
is simultaneously perturbative but such that the phase space density
of gluons is saturated. The saturation scale $\psat$ is estimated,
quantitative numbers for the initial production of gluons 
at the LHC are
given and options for their subsequent evolution are outlined.
For parametrically large nuclei and energies, classical field
methods will be applicable.
\end{abstract}

\section{Perturbative calculability in heavy ion collisions}
The first collisions at RHIC are expected to take place just during this
PANIC conference and the outcome for, say, average charged multiplicity
in a central collision will be known by the time the Proceedings
are out. However, in spite of the fact that QCD is a perfectly well
known theory we cannot with reliability predict this 
outcome -- let alone the properties of QCD plasma produced in
these collisions.

In perturbative QCD reliable predictions can be made for infrared
safe quantities when there is a hard scale $Q$ in the problem,
A prototype is inclusive large $p_T$ jet production, in which the
hard (the parton-parton collision) and the soft (the structure
functions at scale $Q$) parts of the problem can be factorised:

\vspace{-0.5cm}
\begin{center}
\begin{picture}(300,200)(0,0)
\SetWidth{1}
\Line(60,155)(131,155)
\Line(60,145)(131,145)
\Line(60,55)(131,55)
\Line(60,45)(131,45)
\Line(140,150)(210,110)
\Line(140,50)(210,90)
\Line(210,110)(260,150)
\Line(210,90)(260,50)
\Line(210,110)(210,90)
\GCirc(140,150){10}{0}
\GCirc(140,50){10}{0}
\Line(140,150)(170,165)
\Line(140,150)(170,160)
\Line(140,150)(170,155)
\Line(140,50)(170,45)
\Line(140,50)(170,40)
\Line(140,50)(170,35)
\Text(130,170)[lb]{soft}
\Text(270,150)[lb]{$p_T$}
\Text(270,130)[lb]{hard}
\Text(175,155)[lb]{X}
\Text(175,35)[lb]{X}
\Text(40,145)[lb]{A}
\Text(40,45)[lb]{A}
\end{picture}
\end{center}

This schematic diagram is explicitly drawn for a nucleus-nucleus collision,
though calorimetric jet measurements will need very large $E_T$ 
due to the immense
hadronic $E_T$ background. Note also that the unobserved $X$'s will
in an average event contain a huge number of further $2\to2$
collisions.

Consider now the situation as a function of $p_T$ (the scale is for LHC):

\vspace{-1cm}
\begin{center}
\begin{picture}(400,100)(0,0)
\SetWidth{1}
\LongArrow(0,50)(400,50)
\Line(0,50)(0,55)
\Line(50,50)(50,55)
\Line(100,50)(100,55)
\Line(350,50)(350,55)
\Text(0,60)[lb]{0}
\Text(50,60)[lb]{1}
\Text(100,60)[lb]{2}
\Text(330,60)[lb]{100 GeV}
\Text(405,50)[lc]{$p_T$}
\Photon(100,45)(100,15){5}{4}
\Text(95,5)[lb]{$p_\rmi{sat}$}
\Text(10,40)[lc]{Classical field}
\Text(10,25)[lc]{equations}
\Text(110,40)[lc]{Perturbation theory,}
\Text(110,25)[lc]{minijets}
\Text(320,40)[lc]{$\sigma(p_T)$ small,}
\Text(320,25)[lc]{error $\sim 10$\%}
\end{picture}
\end{center}

When one reduces the magnitude of $p_T$, the cross section grows and
the error likewise. In this perturbative minijet picture one can
compute the number of gluons with $p_T\ge p_0$ and the $E_T$ they
carry \cite{bm}-\cite{ek}.
We can continue until $p_0$ reaches the saturation
limit $p_\rmi{sat}$, to be defined below.
Then the gluonic subsystem becomes very
dense and new physics enters \cite{glr}-\cite{mq}. 
For $p_T\ll p_\rmi{psat}$ the phase space occupation number 
becomes $\gg1$. Then 
this new physics may be effectively
described by classical Yang-Mills equations
\cite{elm}, the initial conditions of which are given
by an ensemble in colour space \cite{mv}-\cite{kmw}. 

It is now crucial that for large nuclei and large energies the
gluonic subsystem may become so dense that $\psat$ is in the
perturbative region, $\gsim2$ GeV. 

Numbers for $p_T\gg \psat$ can easily be produced using standard
perturbative techniques. Classical field ideas are not yet
sufficiently developed to permit the same for $\Lambda_\rmi{QCD}\ll
p_T\ll\psat$ and may ultimately require parametrically large
nuclei ($A^{1/3}\gg1$) and energies. To have some quidance from
theory let us use the assumption already advocated in \cite{glr}:
estimate the average event by taking $p_T=\psat$. We are mainly
interested in the local energy density or $E_T$ (per unit
rapidity): gluons with $p_T\gg \psat$ carry lots of $E_T$ but
are very rare, whereas gluons with $p_T\ll\psat$ are numerous
but carry little $E_T$.

\section{Minijet approach to the creation of early little bang}
During the pre-little bang era we have two Lorentz contracted
nuclei approaching each other. The system has zero temperature and
zero entropy. The Lorentz contracted diameters, applicable to the
valence quark parts of the wave functions,
are 
\ba
{2R_A\over\gamma}&=& 1\, \fm\quad\quad\,\,\,\, {\rm SPS},\\ \nn
&=&0.1\,\fm \quad\,\,\,\,\,{\rm RHIC},\\ \nn
&=&0.005\,\fm\quad{\rm LHC}.\nn
\ea
Already this naive fact emphasises the great qualitative
difference between the energy domains. The little bang universe
now is created by the nuclear collision, treated 
in the standard picture as a collision
of two clouds of gluons (which dominate over quarks and antiquarks)
with longitudinal momentum distributions $\approx Axg(x,p_T^2)$.
The entropy content of this universe can be bounded from below by
computing the number of gluons with $p_T>p_0$ = 2 GeV. The dominant
longitudinal momentum fractions are (when gluons from the two
nuclei have equal $p_L=p_T$)
\ba
x &\sim& {2\over 10}\,\,\,\,\,=0.2\quad\,\,\,\,{\rm SPS},\nonumber\\
  &\sim& {2\over 100}\,\,=0.02\quad\,\,{\rm RHIC},\nonumber\\
  &\sim& {2\over 2750}\approx 0.001\,\,\,\,{\rm LHC}.\nonumber
\ea
This is a second naive way of seeing the importance of increasing
energy (keeping $p_0$ fixed): LHC can fully profit from the
enhancement of gluons observed at small $x$ \cite{HERA}.

The computation of the number $N_{AA}$ of gluons produced in a central
A+A collision is quite straightworward \cite{ek}. What is important for
a qualitative discussion is that this number scales $\sim A^{4/3}$.
A hard cross section scales $\sim A^2$, but since one wants the number
per inelastic collision, this has to be divided by $\pi R_A^2\sim
A^{2/3}$:
\be
N_{AA}={A^2\sigma(p_T>p_0)\over \pi R_A^2}\sim A^{4/3}{1\over p_0^2},
\ee
where $\sigma$ is a $2\to2$ cross section. Computed values 
per unit rapidity at $y=0$ are as follows
(shadowing is included, $K=1.5$):

\begin{figure}[h]

\vspace{3.7cm}
\hspace{0.5cm}
\epsfysize=10cm
\centerline{\epsffile{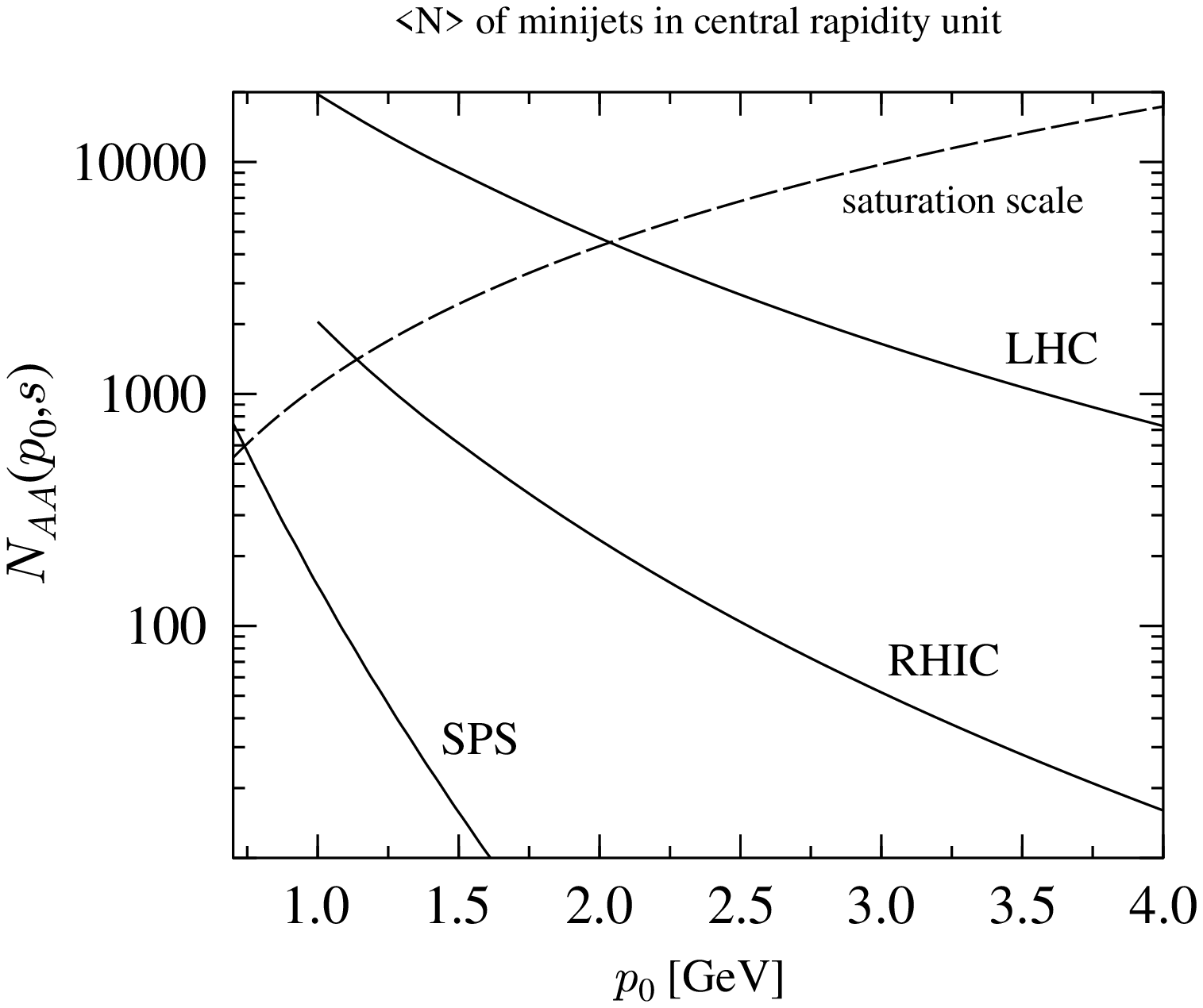}}
\vspace{-8.9cm}

\end{figure}

The result is reliable as long as $p_0$ is large enough so that
perturbation theory is valid and as long as the independent
collision picture can be assumed to be valid. The latter becomes
questionable when $p_0<\psat$, the saturation limit. To define $\psat$,
consider the transverse projection of the nucleus:

\vspace{-1cm}
\begin{center}
\begin{picture}(200,200)(0,0)
\SetWidth{1}
\CArc(100,100)(80,0,360)
\CArc(160,100)(15,0,360)
\LongArrow(190,120)(170,105)
\Text(195,120)[lb]{area of gluon}
\Text(195,100)[lb]{= $\pi/p_T^2$}
\CArc(90,140)(30,0,360)
\LongArrow(30,145)(58,142)
\Text(-20,144)[lb]{Small $p_T$}
\CArc(60,80)(6,0,360)
\LongArrow(26,84)(50,80)
\Text(-20,84)[lb]{Large $p_T$}
\end{picture}
\end{center}

\vspace{-1cm}
One can estimate the each gluon occupies the transverse area
$\pi/p_T^2$ so that the gluon density saturates if
\be
N_{AA}\times\pi/\psat^2 > \pi R_A^2
\ee
from which it follows that $\psat\sim A^{1/6}$. This saturation limit is
also shown in Fig.\ref{naa} and one sees that
\ba
\psat&=& 0.65\, \gev\quad\,\,\,\, {\rm SPS},\\ \nn
&=&1.1\,\gev \quad\,\,\,\,\,{\rm RHIC},\\ \nn
&=&2.0\,\gev\quad\quad{\rm LHC}.\nn
\ea
Extending the computation to include partons with $p_T>\psat$ one has,
in one unit of rapidity:

SPS: 600 gluons with $p_T>0.65$ GeV,\\

\vspace{-0.4cm}
RHIC: 1100 gluons + 110 $q$ + 80 $\bar q$ with $p_T>1.1$ GeV,\\

\vspace{-0.4cm}
LHC: 4300 gluons + 200 $q$ + 190 $\bar q$ with $p_T>2.0$ GeV.

The corresponding $E_T$-values are
\ba
E_T&=& 400\, \gev\quad\quad\quad {\rm SPS},\\ \nn
&=&2500\,\gev \quad\,\,\,\,\,\,\,\,{\rm RHIC},\\ \nn
&=&12000\,\gev\quad\quad{\rm LHC}.\nn
\ea
Remember that these are the {\em initial} values at
$\tau=\tau_i=1/\psat$.

It is, of course, questionable to extend the computation to the
1 GeV region, but the numbers are not unreasonable. However, due
to the (relatively) small Lorentz contraction at SPS it is not
clear how to convert to coordinate space there. At LHC the clocks
can be started very accurately and one can say that
at the LHC there will be at the time 
$\tau_i=1/\psat=0.1$ fm after the collision about 4000 gluons
with $p_T>\psat$ in one unit of rapidity:

\begin{figure}[thb]

\vspace{-0.5cm}
\hspace{0.5cm}
\epsfysize=5cm
\centerline{\epsffile{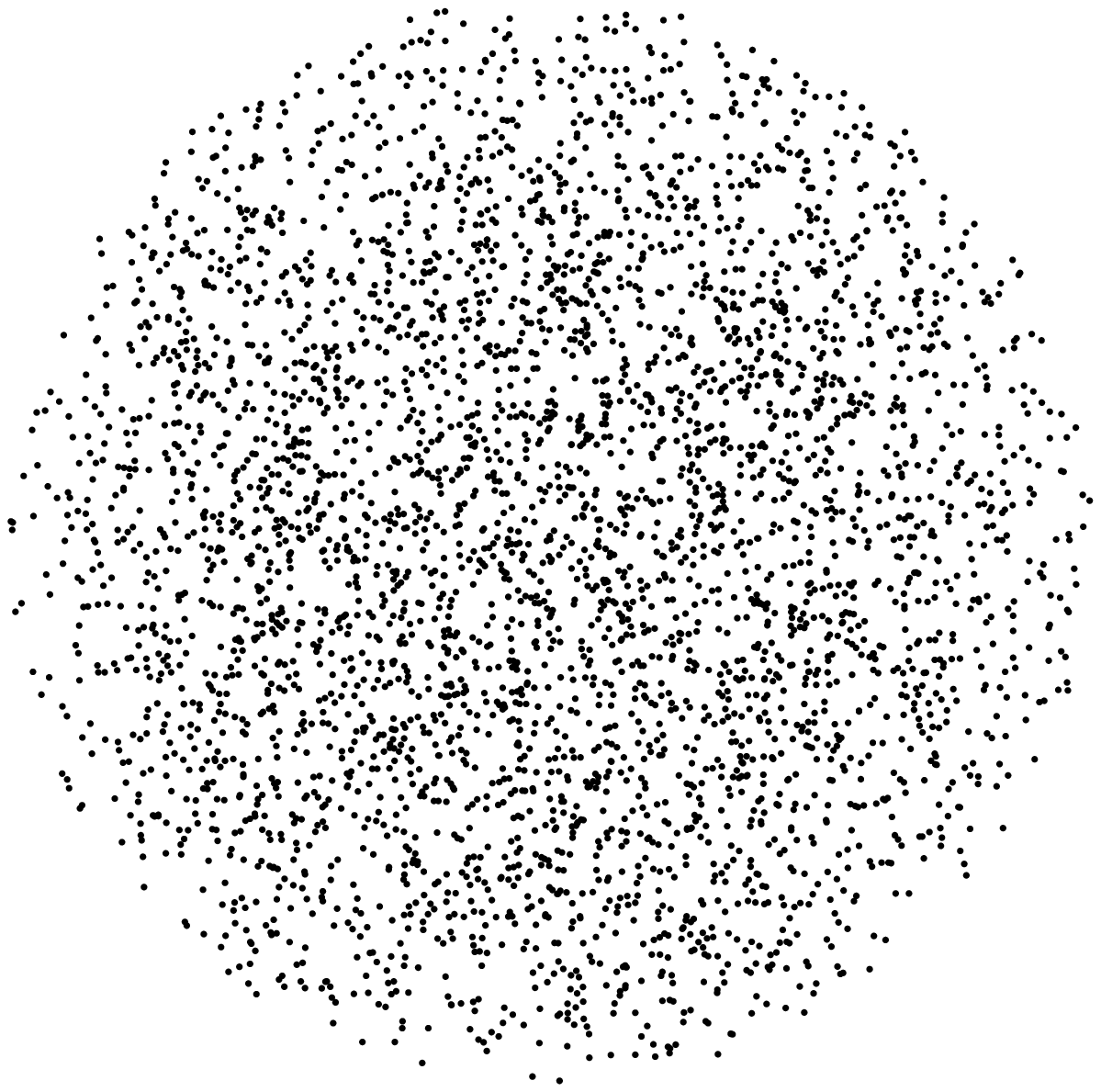}}
\vspace{-1cm}
\label{nonuniform}
\end{figure}

Geometrically, the volume of the little bang universe at the time of
its creation is $V=\pi R_A^2\star \tau_i\Delta y=\pi R_A^2/\psat
= 13\, \fm^3$ for $A$ = Pb. 
This implies that the initial energy and number densities of gluons 
at the LHC are
\be
\epsilon_i=970{\gev\over\fm^3},\qquad n_i={320\over\fm^3}.
\ee

\section{Hubble expansion of the early little bang}
Now comes the crucial question: what then? We have in usual QCD
perturbation theory computed the composition of the $p_T>\psat$ = 2 GeV
component of the system at the time $1/\psat$ = 0.1 fm. Invoking
the assumption of \cite{glr}, this is 
the initial state of an average event. But this is
not yet something comparable with experiment.
How does this subset behave for $\tau>1/\psat$ until conversion
to hadrons?
Of course, nobody knows definitely until a systematic 
non-perturbative computation
in QCD has been carried out.
Thus one has to make assumptions and these will be tested
by experiment.

In this case it is quite natural to
assume that the little bang universe is thermalised at creation at
$\tau=0.1$ fm (note that chemically (anti)quarks are not in equilibrium).
Then it expands conserving the total entropy $S=3.60N\approx15000$.
If we go to the center, the expansion is purely longitudinal,
$S=sV\sim s\tau\sim T^3\tau$ = constant, $s\sim n\sim 1/\tau$ and
$T\sim 1/\tau^{1/3}$ (compare $T\sim 1/t^{1/2} (\sim 1/t^{2/3})$ for
the usual radiation (matter) dominated universe). However, due to
$p\,dV$ work, energy density decreases faster \cite{zgp}:
\be
\epsilon\sim{1\over\tau^{4/3}}\sim n{1\over\tau^{1/3}},\quad
{\epsilon\over n}\sim T\sim{1\over \tau^{1/3}}.
\ee

\begin{figure}[h]

\vspace{3.6cm}
\hspace{0.5cm}
\epsfysize=10cm
\centerline{\epsffile{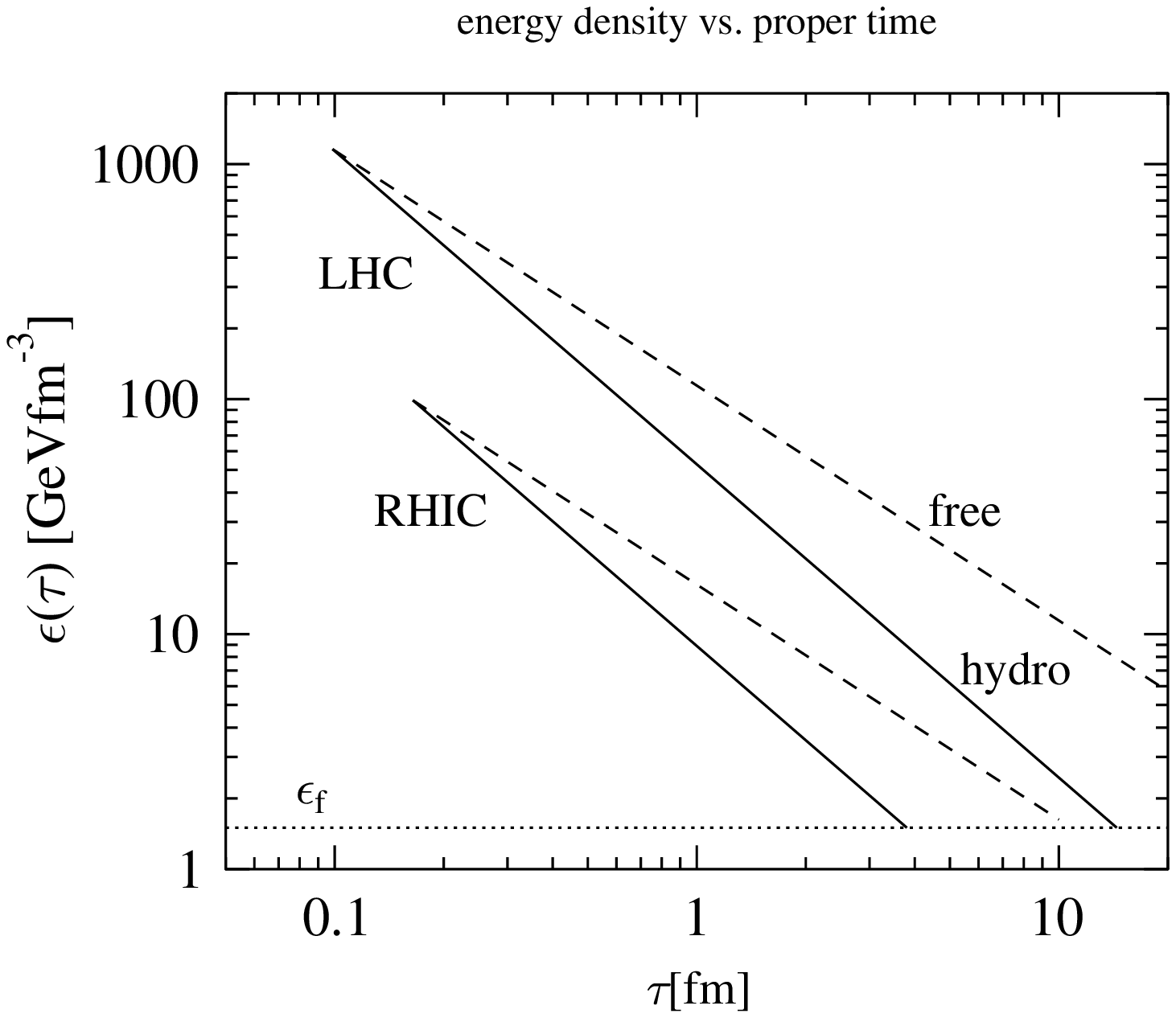}}
\vspace{-8.7cm}
\label{evol}
\end{figure}

Thus the initial and final entropies are the same. The final particles
are dominantly pions, for which also $S\approx4N$. Thus, simply,
$N_\rmi{gluons}=N_\rmi{pions}$! $E_T$ behaves quite differently:
\be
E_{Tf}\lsim\left({\tau_i\over\tau_f}\right)^{1/3}E_{Ti}\sim
\fr16 E_{Ti}.
\ee
Thus the $E_T$ goes down from 12000 GeV to about 2000 GeV. 
This plasma really does work! Stated in another way: the energy per
particle went down from about 3 GeV to about 0.5 GeV. These predictions
for Pb+Pb at LHC can easily be extended to any $A$ and $s$.

Any dissipative effects will increase both the multiplicity and 
$E_T$. The extreme is free streaming: $E_T$ constant. In this limit the
multiplicity would be gigantic, about $E_T/0.5\gev$. If one could
predict the initial parameters with a 10\% accuracy, the measurement
of final quantities would be a measurement of the degree of
thermalisation. Or, if one knew that the expansion is adiabatic,
they would be a measurement of the initial conditions \cite{gmatsui}.

\section{Classical fields}
In the minijet computation above we basically used the $2\to2$ diagram
(x denotes a large $p_T$ external line):

\vspace{-1.5cm}
\begin{center}
\begin{picture}(200,100)(0,0)
\SetWidth{1}
\Line(0,0)(40,0)
\Line(0,60)(40,60)
\Photon(20,0)(20,20){1}{4}
\Photon(20,20)(20,40){1}{4}
\Photon(20,40)(20,60){1}{4}
\Photon(20,20)(40,20){1}{4}
\Photon(20,40)(40,40){1}{4}
\Text(40,20)[]{x}
\Text(40,40)[]{x}
\Text(60,30)[]{=}
\Line(80,30)(120,30)
\Line(120,30)(160,34)
\Line(160,30)(200,30)
\Line(120,26)(160,30)
\Photon(120,30)(137,30){1}{4}
\Photon(137,30)(137,50){1}{4}
\Photon(137,30)(143,30){1}{2}
\Photon(143,30)(143,10){1}{4}
\Photon(143,30)(160,30){1}{4}
\Text(137,50)[]{x}
\Text(143,10)[]{x}
\end{picture}
\end{center}
but what about bremsstrahlung diagrams, like

\vspace{-1.5cm}
\begin{center}
\begin{picture}(200,100)(0,0)
\SetWidth{1}
\Line(0,0)(40,0)
\Line(0,60)(40,60)
\Photon(20,0)(20,30){1}{4}
\Photon(20,30)(40,30){1}{4}
\Photon(20,30)(20,60){1}{4}
\Text(40,0)[]{x}
\Text(40,30)[]{x}
\Text(60,30)[]{=}
\Line(80,30)(120,30)
\Line(120,30)(160,34)
\Line(160,30)(200,30)
\Line(160,30)(110,10)
\Photon(120,30)(150,30){1}{4}
\Photon(150,30)(150,50){1}{4}
\Photon(150,30)(160,30){1}{4}
\Text(150,50)[]{x}
\Text(110,10)[]{x}
\Text(220,30)[]{?}
\end{picture}
\end{center}
This diagram predicts \cite{bg} a certain number of gluons around
$y=0$, but how to extend this computation to nuclear collisions?
This is not as straightforward as in the minijet case: the $p_T$ of
the perturbative gluon is balanced by a ``beam jet''.  The new ideas
developed \cite{mv,kmw}, \cite{kr}-\cite{kv} amount to computing classical
radiation of gluons by two colour currents, formed by the valence
quarks and the fastest gluons of the colliding nuclei, on the forward
and backward light cones.

The motivation for the use of the classical field approximation is
precisely that in the saturation limit the occupation number of any
of the field modes is large and the classical field approximation should
be appropriate. The situation is analoguous to that of computing the
rate (``the sphaleron rate'')
of the baryon number violating reactions in electroweak matter
\cite{moore}. These reactions also involve nonperturbative large
field modes and their rate can only be computed numerically.

The essential parts of the classical field computation are the 
formulation of the equations of motion (``effective theory'') and
of the initial conditions. In the electroweak case, the standard model
at finite $T$, the theory is weakly coupled and there is a perfectly
controllable perturbative way \cite{klrs} of computing the
effective theory, which is a three-dimensional
purely Euclidian SU(2)$\times$U(1) gauge + fundamental (doublet) 
Higgs theory. The initial
conditions are set by thermalising the system at some temperature, which
is numerically straightforward. After that the classical field equations
are integrated numerically in real time
starting from some thermal field configuration and 
the sphaleron rate is
computed as a real time correlation function of the topological
susceptibility. 
Hereby one has learnt that
the dynamics of the infrared small-momentum modes one is interested
in is essentially affected by their interactions with the hard
large-momentum ($\sim T$) modes. When this is correctly taken into
account, the calculation of the sphaleron rate, which has no perturbative
contribution, should be under control \cite{moorerummu}.
This is not yet the case with
quantities like viscosities, which also have a perturbative contribution.

The situation is much more complicated in the application
of the classical field approach to QCD in heavy ion collisions. 
Here there is no ``top down'' systematic way of deriving the effective
theory, even in a theoretical weakly coupled limit and one has to
proceed phenomenologically \cite{mv}. Let
us focus our attention at the slice of matter at rest in the CMS.
Then our space is again three-dimensional, but in this case
Minkowskian with time and two spatial coordinates. The effective theory
now is an SU(3) gauge + adjoint (octet) Higgs theory \cite{kv},
but there is no known systematic way to derive its coefficients. 
To see why the theory is of this form, take
the gluon part of the QCD action and
include only longitudinal boost invariant configurations,
independent of $\eta=\tanh^{-1}(z/t)$, which implies
an infinite energy. The action then becomes
\be
S[A_\mu^a,\phi^a]=\int d\tau\,d^2x
\left[-{\tau\over4}F_{\mu\nu}^aF^{a\mu\nu}+
{1\over2\tau}(D_\mu\phi)^a(D^\mu\phi)^a\right],
\ee
where $\mu=0,1,2$, $A^a_1,A^a_2$ are the transverse components of the
gluon field, $\phi^a$ is the component $A^a_\eta$ of the gluon field
and the gauge has been chosen as $A^a_\tau=0$. This part of the action is
singular at $\tau=0$, on the light cone. This singularity is
regulated by the dynamics of the nuclear interaction at $\tau=0$,
by the initial condition. One collision is represented by one
random charge distribution $\rho^a_{(i)}(x_1,x_2)$ 
of the two colliding nuclei ($i=1,2$), where each of the two $\rho$'s
is drawn from a, say, Gaussian distribution of some width $\mu$.
The $\rho$'s give the initial conditions for the fields 
$A_\mu^a(0,x_1,x_2),\phi^a(0,x_1,x_2)$ and the equations
of motion give their further evolution in $\tau$. Various physical
quantities, like the number of bremsstrahlung gluons, 
can then be computed by
averaging over many different evolutions corresponding
to different initial conditions drawn from the Gaussian ensemble.
A first attempt to carry out this program (for SU(2)) is in \cite{kv}.

The crucial parameter
is the average transverse colour charge density $\mu^2$, which for the
valence quarks of a nucleus is given by
\be
\mu^2= {(N_c^2-1)\over2N_c^2}{A\over\pi R_A^2},
\ee
but to which also the (dominating) 
contribution of the fast gluons has to be
added \cite{gm}. The computation then is formally valid
for
\be
\Lambda_\rmi{QCD}\ll p_T\ll\mu,
\ee
where the lower limit expresses the fact that one is in the
weakly coupled domain ($\alpha_s$ is parametrically small) and
the upper that one simultaneously is in the saturation domain,
occupation numbers are large.
Parametrically, like $\psat$, $\mu$ scales $\sim A^{1/6}$.
However, 
numerically at LHC energies, $\mu\approx0.8$ GeV \cite{gm}, while
$\psat$ was about 2.0 GeV. Converted to time units this means that
at LHC the bremsstrahlung quanta thus will materialise at times
$0.25\,\fm\,\lsim\,\tau\,\lsim\,1\,\fm$, while the perturbative quanta
materialised at $\tau\approx0.1\fm$. For practical applications,
the window of applicability of the classical
field approach thus seems very small, but conceptually it deals with an
entirely new domain of QCD and thus studies with parametrically large
$A,1/g$ and $\sqrt{s}$ are very important.

However, even if one could solve the problem for all the scales
$p_T\gg\Lambda_\rmi{QCD}$, implying the computation of the initial
creation of the system at all times $\ll$ 1 fm, there still
remains the problem of all smaller $p_T$-scales, the entire
further evolution of the system, its expansion, conversion to
the hadronic phase and ultimate decoupling. Here one in any case
will have to resort to phenomenological ideas, which later will
be checked by experiments.

\section{Conclusions}
Heavy ion experiments at RHIC and LHC will enter a new domain of QCD,
in which the production of even large $p_T$ gluons is so 
abundant that it can be simultaneously perturbative but such that
that the phase space density of gluons is saturated. This talk has
discussed precisely what numbers of gluons can be expected under
these conditions to be initially (at 0.1 fm) produced
at the LHC and what options there are for the subsequent
behaviour of this system. Whatever the detailed validity of these
numbers is, one can in any case with confidence look forward to
abundant production of QCD plasma at RHIC and especially at the
LHC.


\begin{thebibliography}{9}

\bibitem{bm}
J.P.~Blaizot and A.H.~Mueller,
Nucl. Phys. {\bf B289} (1987) 847.

\bibitem{kll}
K.~Kajantie, P.V.~Landshoff and J.~Lindfors,
Phys. Rev. Lett. {\bf 59} (1987) 2527.

\bibitem{ekl}
K.J.~Eskola, K.~Kajantie and J.~Lindfors,
Nucl. Phys. {\bf B323} (1989) 37.

\bibitem{ek}
K.J.~Eskola and K.~Kajantie,
Z. Phys. {\bf C75} (1997) 515,
nucl-th/9610015.

\bibitem{glr}
L.V.~Gribov, E.M.~Levin and M.G.~Ryskin,
Phys. Rept. {\bf 100} (1983) 1.

\bibitem{mq}
A.H.~Mueller and J.~Qiu,
Nucl. Phys. {\bf B268} (1986) 427.

\bibitem{elm}
H.~Ehtamo, J.~Lindfors and L.~McLerran,
Z. Phys. {\bf C18} (1983) 341.

\bibitem{mv}
L.~McLerran and R.~Venugopalan,
Phys. Rev. {\bf D49} (1994) 2233,
hep-ph/9309289.

\bibitem{kmw}
A.~Kovner, L.~McLerran and H.~Weigert,
Phys. Rev. {\bf D52} (1995) 6231,
hep-ph/9502289.

\bibitem{HERA}
N.H.~Brook
[for ZEUS \& H1 Collaborations],
hep-ex/9805031.

\bibitem{zgp}
B.~Zhang, M.~Gyulassy and Y.~Pang,
Phys. Rev. {\bf C58} (1998) 1175,
nucl-th/9801037.

\bibitem{gmatsui}
M. Gyulassy and T. Matsui, 
Phys. Rev. {\bf D29} (1983) 419.

\bibitem{bg}
J.F.~Gunion and G.~Bertsch,
Phys. Rev. {\bf D25} (1982) 746.

\bibitem{kr}
Y.V.~Kovchegov and D.H.~Rischke,
Phys. Rev. {\bf C56} (1997) 1084,
hep-ph/9704201.

\bibitem{gm}
M.~Gyulassy and L.~McLerran,
Phys. Rev. {\bf C56} (1997) 2219,
nucl-th/9704034.

\bibitem{mmr}
S.G.~Matinian, B.~M\"uller and D.H.~Rischke,
Phys. Rev. {\bf C56} (1997) 2191,
nucl-th/9705024.

\bibitem{kv}
A.~Krasnitz and R.~Venugopalan,
hep-ph/9809433.

\bibitem{moore}
G.D.~Moore,
hep-lat/9907009.

\bibitem{klrs} K. Kajantie, M. Laine, K. Rummukainen 
and M. Shaposhnikov,
Nucl. Phys. {\bf B458} (1996) 90,
hep-ph/9508379.

\bibitem{moorerummu}
G.D.~Moore and K.~Rummukainen,
hep-ph/9906259.

\end{thebibliography}
\end{document}